# Presentation Attack Detection using Convolutional Neural Networks and Local Binary Patterns

Justin Spencer
Undergraduate Student
*Computer Science Department*
*North Carolina Agricultural and Technical State University*
Greensboro, North Carolina, United States of America
bjspencer@aggies.ncat.edu

Deborah Lawrence
Undergraduate Student
*Computer Science Department*
*University of North Carolina-Asheville*
Ashevile, North Carolina, United States of America
dlawrenc@unca.edu

Prosenjit Chatterjee
Doctoral Student
*Computer Science Department*
*North Carolina Agirucultural and Technical State University*
Greensboro, North Carolina, United States of America
pchatterjee@aggies.ncat.edu

Dr. Kaushik Roy
Associate Professor
*Computer Science Department*
*North Carolina Agricultural and Technical State University*
Greensboro, North Carolina, United States of America
kroy@ncat.edu

Dr. Albert Esterline
Associate Professor
*Computer Science Department*
*North Carolina Agricultural and Technical State University*
Greensboro, North Carolina, United States of America
esterlin@ncat.edu

Dr. Jung-Hee Kim
Assistant Professor
*Computer Science Department*
*North Carolina Agricultural and Technical State University*
Greensboro, North Carolina, United States of America
jungkim@ncat.edu

*Abstract*—The use of biometrics to authenticate users and control access to secure areas has become extremely popular in recent years, and biometric access control systems are frequently used by both governments and private corporations. However, these systems may represent risks to security when deployed without considering the possibility of biometric presentation attacks (also known as spoofing). Presentation attacks are a serious threat because they do not require significant time, expense, or skill to carry out while remaining effective against many biometric systems in use today.

This research compares three different software-based methods for facial and iris presentation attack detection in images. The first method uses Inception-v3, a pre-trained deep Convolutional Neural Network (CNN) made by Google for the ImageNet challenge, which is retrained for this problem. The second uses a shallow CNN based on a modified Spoofnet architecture, which is trained normally. The third is a texture-based method using Local Binary Patterns (LBP). The datasets used are the ATVS-FIr dataset, which contains real and fake iris images, and the CASIA Face Anti-Spoofing Dataset, which contains real images as well as warped photo, cut photo, and video replay presentation attacks. We also present a third set of results, based on cropped versions of the CASIA images.

## I. INTRODUCTION

Researchers have studied ways to measure and differentiate between different palm prints, gaits, voices, fingerprints, irises, faces, and other biometric identifiers. All are interesting and effective ways to verify an identity, although this research only examines iris and facial biometrics. Through a variety of methods, it has become possible for a computer system to scan and analyze a face or iris and then grant access based on whether that biometric is recognized.

Biometric-based presentation attacks involve gaining access to a biometric sample from databases or external resources, then reusing that biometric to gain unauthorized access to confidential data or secure facilities. Though the use of biometric authentication strengthens security through unique features, the cloning of the biometric sample and those unique features to access a biometric system illegally is feasible.

Though face and iris recognition are more reliable biometrics, spoofing has still become a common threat for these biometrics. There are multiple ways an attacker can spoof a biometric system. High-resolution copies of biometric samples have been used to spoof systems. Photorealistic face masks and synthetic images have also been used successfully in presentation attacks. Digital retouching of images is also a common spoofing threat [1]. Face and iris spoofing can be categorized as texture-based spoofing, motion-based spoofing, 3D shape-based spoofing, and multi-spectral reflectance-based spoofing [2].

## II. RELATED WORK

We investigate several forms of presentation attack detection that use machine-learning methods, including Convolutional Neural Networks (CNNs) and deep belief networks. We also examine techniques for the more general problem of facial recognition and classification that are useful for our research.

This research is based upon work supported by the National Science Foundation (grant number CNS-1460864) and the Army Research Office (contract number W911NF-15-1-0524).

Deep learning techniques to mitigate presentation attacks have shown promising results [1, 2, 5]. Menotti et al. proposed building an anti-spoofing system using a CNN with a combination of two approaches [5]. The first approach focuses on learning an appropriate CNN architecture. The second approach consists of learning filter weights via the standard backpropagation algorithm [9]. Yang, Lei, and Li propose techniques including face localization, spatial augmentation, and temporal augmentation in combination with canonical CNN filtering techniques for feature training and classification [2]. CNNs are the primary focus of this research.

Silva et al. use a deep belief network in order to detect whether a user is wearing contact lenses [12]. They define a three-class detection problem by dividing the images based on the presence of soft (uncolored) contact lenses, colored contact lenses, and no contact lenses. They use a combination of a CNN for deep image representations and a fully connected three-layer network for classification. Instead of using a specific search algorithm, the researchers analyzed a set of parameters to build the final network topology and to learn the filter weights by backpropagation. They have also suggested future work with random weights. On certain databases, their methods outperform state-of-the-art approaches. However, their current approach does not segment the iris, and this becomes a problem in datasets where the iris region is not pre-identified.

Farfade, Saberian, and Li investigate facial recognition for partially obscured faces and faces at an angle [13]. Using a specific CNN called Deep Dense Face Detector (DDFD), the authors create a system that can recognize facial features without the entire face being visible. They specifically were interested in creating a single model that could recognize faces despite a variety of obstructions, such as rotated faces, skewed faces or faces in profile, partially obscured faces, and others. DDFD does not require landmark annotation, meaning it does not specifically pick out images of eyes, noses, and other significant facial features. Instead, the authors used a dataset of 21,000 unaltered and altered images featuring partially obscured human faces in order to train the CNN. Altered images involved sampling the dataset and randomly cropping certain images in order to represent an obscured face. By training the CNN to recognize cropped images, it could recognize faces in a variety of situations. DDFD used five convolutional and three fully connected layers, and produced classification results that were comparable with state-of-the-art results in other works.

Garcia and Delakis focus primarily on facial recognition with images that have been captured in non-controlled environments [14]. These images are of variable size, quality, and rotation. The authors used a dataset created from images from the internet and scanned newspapers to create a CNN called the Convolutional Face Finder (CFF) specifically for their work. When processing images, CFF classifies various local features such as end-points or edges, combines them in later layers to identify larger features such as noses or eyes, and then measures the feature's distance relative to other features to recognize a face.

Yang, Lei, and Li are perhaps the first to suggest using a CNN to differentiate between real and spoofed faces [2]. They crop the CASIA and Replay-Attack database images to five different sizes in order to determine the influence of the background in presentation attack detection. Their best result is with the second-largest size, possibly due to the largest size resulting in overfitting. Their research shows that the background can be useful in certain image identification problems, and that CNNs are an effective solution for presentation attack detection.

Li et al. also propose the use of a CNN in presentation attack detection [15]. They retrained the VGG-face model created by the Oxford Visual Geometry Group (VGG) on the CASIA and Replay-Attack datasets. Their approach has reasonable performance in comparison with previous work on CASIA and Replay-Attack.

Farfade, Saberian, and Li, as well as Garcia and Delakis, do not directly investigate presentation attack detection. However, their respective work on detecting obscured faces and faces in non-ideal environments is a necessary condition for presentation attack detection. These attacks may be carried out in poor lighting or with partially obscured faces, so any presentation attack detection technique designed for real-world use must be able to operate effectively under these conditions. The research by Silva et al. on contact lens detection is similarly applicable to detecting iris presentation attacks that use textured contact lenses. Finally, the research by Menotti et al., Yang, Lei, and Li, and Li et al. helped inform our decision to use CNNs and to test cropped versions of the CASIA images.

## III. THEORETICAL BACKGROUND

Presentation attack detection algorithms are classified as either hardware-based or software-based methods. We investigate software-based methods, which are cheaper since they do not require specific hardware and tend to be more user-friendly since they do not require a challenge-response [16]. Software-based methods can be divided into additional subclasses, such as dynamic methods that use temporal information (e.g., videos) or static methods that do not (e.g., still images). We use only software-based static methods for still images. Techniques examined include a texture-based approach using Local Binary Patterns (LBP) and a machine learning-based approach using CNNs.

Texture-based approaches in image analysis use microtextural data in order to determine characteristics of the images provided. LBP is a texture classification technique that compares a single pixel with its eight neighboring pixels [17]. Moving in a circle around the center pixel, LBP compares the brightness of the center pixel with each individual neighbor, determining whether the neighbor has a larger (brighter) or smaller (darker) value than the center pixel. If the neighbor is brighter than the center cell, it is assigned a value of one. If it is equally bright or darker than the center pixel, it is assigned a value of zero. These values are then converted into a histogram. After performing this process for a series of pixels, the histograms are combined to produce a feature vector. Finally, this feature vector is used to classify the image. LBP has proven to be effective in image classification and facial recognition, and can be applied to presentation attack detection due to microtextural differences between real human faces and printed photos or iPad screens used for spoofing [18].

CNNs are a heavily modified form of traditional neural networks designed for image recognition. Traditional neural networks are made out of a number of different layers. Each neuron in a layer receives input signals from all the previous layer's neurons, and transmits an output signal to all the next layer's neurons if and only if the input signal strengths exceed some threshold. Because each neuron in a layer is fully connected to every neuron in the previous layer, the number of connections grows rapidly with the number of layers in the network as well as the number of neurons in each layer.

In practice, traditional neural networks support a very limited number of layers, require large amounts of training data even for small networks, and have high hardware requirements and performance costs [4]. Since they do not scale effectively, it is difficult to use them for image recognition and other problems with complex input domains.

CNNs differ from traditional neural networks in that neurons in each layer are not fully connected to all the previous layer's neurons. This is because not every pixel in an image is related to every other – for example, background pixels are not related to pixels of a subject's face [4]. By removing unnecessary connections between neurons, the size and complexity of the network is reduced, allowing CNNs to have more layers and achieve better results on image recognition problems. They also become desensitized to minor variations in the input image. Because of these factors, they are especially efficient tools for facial recognition and classification.

Finally, we explore a technique called transfer learning. Transfer learning involves the reuse of a neural network that has already been trained by other researchers on a different problem. Normal training involves initializing the neuron weights to some value, and then updating them to achieve better accuracies using backpropagation. In transfer learning, a neural net trained on one problem (e.g., the ImageNet challenge) is retrained on a new, similar problem (e.g., presentation attack detection). Since the two problems are similar, the neural network does not need to be fully trained on the new problem, and can reuse what it learned on the original problem. To achieve this reuse of parameters, most of the layers have their weights held constant, and only a few layers have their weights updated. This greatly reduces the time and amount of data required to train the neural network, although the accuracy suffers slightly since the two problems are not exactly the same. For larger CNNs such as Inception-v3, transfer learning can reduce the training time from several weeks to several hours.

## IV. RESEARCH OVERVIEW

Our work will measure the effectiveness of LBP, a shallow CNN, and a deep CNN on cropped and uncropped images. We design and implement a shallow CNN that successfully classifies real and spoofed faces and irises. We then compare the accuracy of our fully trained CNN versus a pre-trained Inception-v3 instance that is retrained for this problem. Inception-v3 is a deep CNN designed by Google and trained on the ImageNet database [19].

Our goal is to examine the accuracy of a shallow CNN architecture versus a deep CNN architecture and determine the ratio between accuracy, performance, and complexity. Our CNN is a modified version of the Spoofnet architecture with six layers whereas the Inception-v3 model has 48. We also compare the effectiveness of ordinary training versus transfer learning for presentation attack detection.

Next, we use LBP to detect presentation attacks, and compare this texture-based method with the above machine-learning methods. LBP was chosen as a comparison technique because it is generally an effective method for classifying multiple kinds of presentation attacks [16]. Including LBP allows us to compare two types of CNNs with an entirely different technique for facial recognition that is still highly accurate [21].

By using both cropped and uncropped images, we intend to compare the accuracy of the different tools when background information is both present and absent. Cropping theoretically should not affect the texture-based methods but will likely affect the CNNs due to the loss of background information. In previous work, it has been shown that CNNs struggle with over-cropped images but perform best when some cropping is done to prevent overfitting [2].

We use multiple datasets with multiple forms of image recognition in order to compare the relative performance of these methods. The results of this work can inform future researchers about the effectiveness of different techniques for presentation attack detection.

## V. CNN ARCHITECTURE

Generally, a CNN architecture is formed by a stack of discrete layers [4]. It transforms the input volume into an output volume through a differentiable function [4]. Inspired by the Spoofnet framework used in [5], we designed our own CNN architecture. The first layer in our CNN is a convolutional layer with a 5x5 kernel, 16 filters, and ReLU activation. Next, there is a max-pooling layer with a 3x3 pooling window and a stride of three. The third layer is convolutional, and is identical to the first except for having 32 filters. This is followed by a second max-pooling layer that is identical to the first. We flatten the output of this layer before feeding it to a dense layer with 128 neurons and ReLU activation. Finally, there is a dense layer with one neuron and sigmoid activation. We use binary cross-entropy as the loss function, a learning rate of 0.001, and the Adam optimizer. Fig. 1 below shows the major features of our CNN architecture.

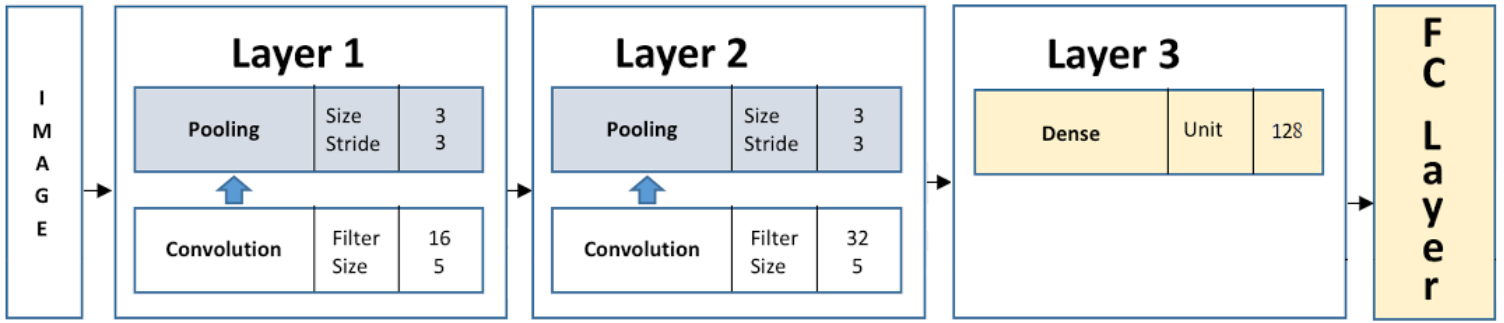

Fig. 1. Diagram of our CNN architecture and layer features.

## VI. DATASETS

We report results for three datasets: ATVS, CASIA, and CASIA-cropped. ATVS is an iris image dataset that contains fifty subjects [22, 23]. Subjects had each eye photographed four times in each of two different sessions. Each image was then printed and rescanned at a reduced quality to create the spoofed iris images. There are 32 images per user, 800 per class (real and fake), for a total of 1600. Each image is a grayscale 640x480 BMP file, which we converted to JPG for use with Inception-v3.

CASIA is a face video dataset containing fifty subjects [11]. Within each class, a subject has one low-resolution landscape-style video, one low-resolution portrait-style video, and one high-resolution portrait-style video that are approximately ten long. The four classes are real subjects, "warped photo" presentation attacks (printed photo of subject held up to the camera, photo is moved back and forth to fool liveness detection systems), "cut photo" attacks (printed photo of subject with eyeholes cut out, real user positioned behind photo to fool blinking detection systems), and video replay attacks (tablet or screen held up to the camera while playing a video of subject).

We converted each MP4 video to a series of still JPGs, with one image per frame of video. Not all videos are the same length, and therefore not every class has the same number of images. Some (but not all) high-resolution videos had an initial black frame, which was discarded. The low-resolution images are 640x480 (landscape style) and 480x640 (portrait style). The high-resolution images are 720x1280 (portrait style). In total, there are approximately 111,000 color images after these transformations, with 20,000 to 30,000 per class.

CASIA-cropped is a custom dataset that we created based on modified versions of the CASIA images. We used pretrained OpenCV Haar cascades [20] to detect and crop the face region for every subject image in the CASIA database. Images without a detected face region, or with more than one detected face region, were discarded. We then cropped the resulting images a second time to ensure all images were the same size (140x140). There are approximately 98,000 color images in this dataset, with 15,000 to 25,000 per class. Cropping the images creates a more challenging problem for the classifier, since a CNN can no longer learn information about the background region to detect spoofing. In the standard CASIA dataset, for example, a CNN may detect presentation attacks by locating the edges of a photo or tablet held up to the camera. In CASIA-cropped, this is no longer possible, and the neural net must learn other, less obvious features to detect spoofing. This helps to create a more robust spoofing detection system that will generalize better to new types of attacks. For example, a photorealistic mask would not have well-defined edges, and might fool systems that use edge detection to classify attacks.

## VII. METHODOLOGY

Inception-v3 was originally trained on the ImageNet dataset, is retrained on these datasets for 4000 epochs, and uses an 80%/10%/10% training/testing/validation split. It does not require that input images be the same size, and loads the full contents of every dataset in batches.

Our custom CNN is trained normally for 30 epochs, uses a 50%/50% training/testing split, and uses the first 30 testing images as a validation set. Unlike Inception-v3, it requires that all input images be the same size. Note that our CNN code can load the full ATVS dataset due to its smaller size but cannot load the larger CASIA and CASIA-cropped datasets. This is because it must load the entire dataset at once, instead of in batches like Inception-v3. We therefore load 800 640x480 images per class for CASIA, and 800 140x140 images per class for CASIA-cropped. The number of images per class, and the image size used for CASIA, were chosen to match ATVS. Although our CNN obtains reasonable accuracy when it converges, it fails to find a gradient about one-third of the time.

We can only generate Receiver Operating Characteristic (ROC) curves for our CNN on binary classification problems. Therefore, we also report accuracies for a simpler binary classification problem using both CASIA datasets. We place all warped photo attack, cut photo attack, and video replay attack images into a single "fake" class, which is compared with the existing "real" class.

Our LBP code also requires that all input images be the same size but is able to load all images of that size instead of only 800 per class. We load all 640x480 images for CASIA and all 140x140 images for CASIA-cropped. We report the smallest number of patches that achieves the highest accuracy.

## VIII. RESULTS AND DISCUSSION

Our results are shown in Table 1 below. The spoofed ATVS images are quite simple and are easily detected by all methods. The CASIA images are more complex, and although our CNN performs well on the regular images, the cropped images force our CNN to guess randomly for the full four-class problem. Binary classification on CASIA-cropped is simpler, and our CNN does much better on that problem due to its reduced dimensionality. LBP and Inception-v3 both perform extremely well regardless of the image size or dataset used.

TABLE I. RESULTS MATRIX FOR INDIVIDUAL TOOLS AND DATASETS.

| Result for dataset (right) and tool (below) | ATVS | CASIA | CASIA-cropped |
|---|---|---|---|
| Inception-v3 | 100% R/F | 98.7% W/C/V/R | 90.2% W/C/V/R |
| Modified Spoofnet | 97% R/F | 90.5% W/C/V/R | 25% W/C/V/R |
|  |  | 92.5% R/F | 94.5% R/F |
| LBP | 100% R/F (1x1 patch) | 100% W/C/V/R (1x1 patch) | 100% W/C/V/R (1x1 patch) |

a. W: Warped photo, C: Cut photo, V: Video presentation attacks. R: Real images. Original CASIA four-class problem.

b. R: Real images. F: Fake images. For CASIA and CASIA-cropped, "Fake" includes all warped photo, cut photo, and video replay attack images).

The following ROC curves show the True Positive Rate (TPR, solid line) versus the False Positive Rate (FPR, dashed line) for our CNN on the ATVS, CASIA, and CASIA-cropped datasets. The TPR should be as close to 1.0 as possible, meaning that all legitimate users were accepted and none were rejected by mistake. The FPR should be as close to 0.0 as possible, meaning that all presentation attacks were rejected and none were mistakenly accepted as real users.

In all cases, the area under the curve remains at 0.98, which is extremely close to the ideal value of 1.0. However, the TPR is generally higher for the simpler ATVS images than for the CASIA-based ones at the same FPR. This indicates better performance, possibly due to the more controlled nature of the real ATVS images or the lower quality of the spoofed ATVS images.

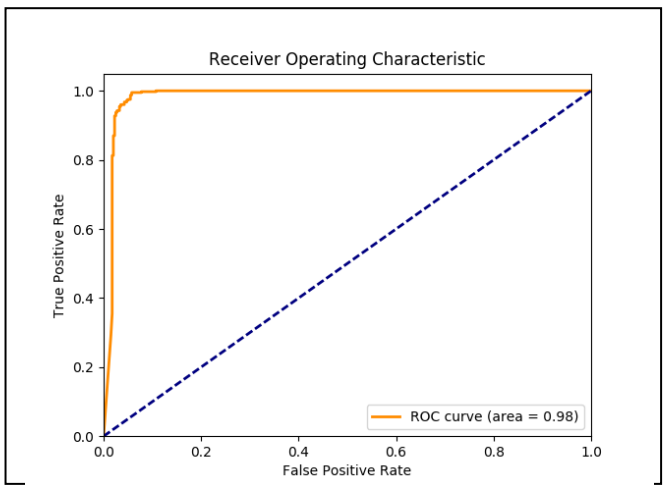

Fig. 2. ROC curve for our CNN on ATVS.

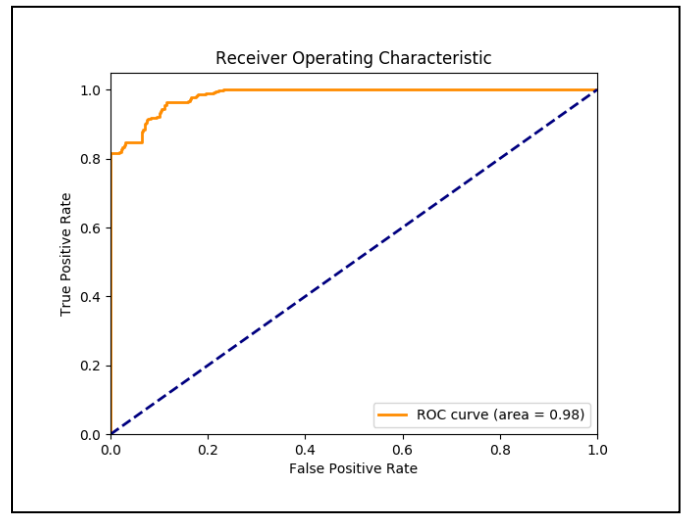

Fig. 3. ROC curve for our CNN on the modified CASIA real/fake problem.

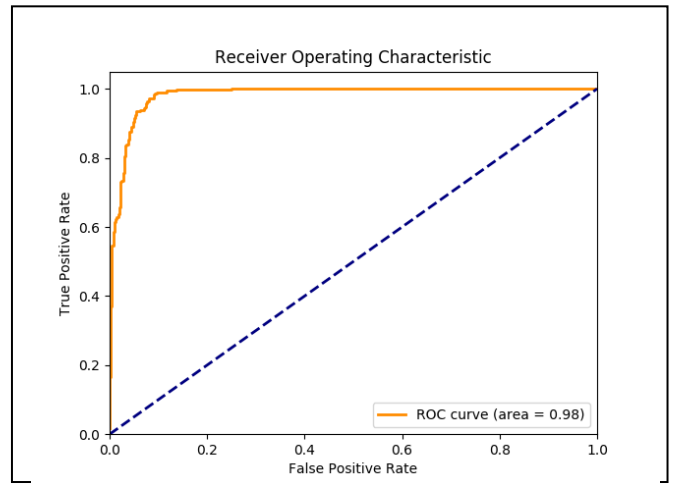

Fig. 4. ROC curve for our CNN on the modified CASIA-cropped real/fake problem.

IX. CONCLUSION

We compare the use of a deep CNN with the use of a shallow CNN for biometric presentation attack detection in still images. We also compare the use of a texture-based method versus machine-learning methods and the effectiveness of transfer learning versus normal training. Finally, we demonstrate a modified Spoofnet architecture that can effectively distinguish between presentation attacks and real users and can classify presentation attacks by type under certain conditions.

For future work, we would like to enhance our CNN code to be more comparable to Inception-v3. These enhancements would include loading data in batches (to make it possible to load all images in the CASIA datasets), loading input images of different sizes, and using the same proportions of training, testing, and validation data as Inception-v3. We would also like to implement some form of data augmentation and multi-class ROC curves.

ACKNOWLEDGMENT

We would like to acknowledge our graduate student mentor John Jenkins, as well as Dr. Joseph Shelton, who provided code to analyze images using LBP, and Ryan Dellana, who provided support for our custom CNN code.